\documentclass[nofootinbib,preprint,aps,draft,pre,superscriptaddress,citeautoscript,floatfix]{revtex4-2}

\usepackage{graphics}
\usepackage{graphicx}
\usepackage{mathrsfs}
\usepackage{amsmath}
\usepackage{bm}
\usepackage{color}
\usepackage{float}
\def\Ray{\mathcal {R} }

\begin{document}

\title{Contact Angle of an Evaporating Droplet of Binary Solution on a Super Wetting Surface}

\author{Mengmeng Wu}
\affiliation{Center of Soft Matter Physics and its Applications, Beihang University, Beijing 100191, China}
\affiliation{School of Physics, Beihang University, Beijing 100191, China}
\author{Masao Doi}
\affiliation{Center of Soft Matter Physics and its Applications, Beihang University, Beijing 100191, China}
\affiliation{School of Physics, Beihang University, Beijing 100191, China}
\author{Xingkun Man}
\email{manxk@buaa.edu.cn}
\affiliation{Center of Soft Matter Physics and its Applications, Beihang University, Beijing 100191, China}
\affiliation{School of Physics, Beihang University, Beijing 100191, China}

\begin{abstract}
We study the dynamics of contact angle of a droplet of
binary solution evaporating on a super wetting surface.  Recent 
experiments show that although equilibrium contact angle of such droplet is zero,
the contract angle can show complex time dependence  before reaching the 
equilibrium value.  We analyse such phenomena by 
extending our previous theory for the dynamics of an
evaporating single component droplet to double component droplet. 
We show that the time dependence of the contact angle can be quite complex. Typically, it
first decreases slightly, and then increases and finally decreases again. Under
certain conditions, we find that the contact angle remains constant over a
certain period of time during evaporation.
We study  how the plateau or peak contact angle depends on the initial composition
and the humidity.  The theory explains experimental 
results reported previously.
\end{abstract}

\maketitle

\section{Introduction} \label{sec.1}

Evaporation of liquid droplets is a phenomenon commonly seen in our daily life.  
It is also important  in recent technologies, such as
ink-jet printing~\cite{Thokchom2017,Kim2012}, nanopatterning
depositions~\cite{Prevo2004,Park2006,Harris2008,Yella2009,Wang2012}, and 
medical diagnostics~\cite{Yakhno2005,Carreon2018}.

Evaporation of liquid droplets is a material transport problem in which 
the capillary flow of liquid is coupled with phase change and diffusion, 
and includes surprisingly complex problems of basic science. 
The complexity is seen in the simple case that a single component droplet is evaporating
on a super-wetting surface. If the liquid is non-volatile and not evaporating, the
droplet spreads on the surface, but if the liquid is evaporating, the
droplet first spreads, but then recedes due to evaporation. It has been
reported that in such situation, the contact angle  often shows 
quasi-stationary, apparent contact angle $\theta_{app}$.  
The phenomena has been studied extensively 
by experiments and simulation \cite{Picknett1977,
BourgesMonnier1995,Erbil2002,Stauber2014,Stauber2015,Shrikanth2019}. 

The evaporation behavior of droplets becomes even more complex 
for solutions~\cite{Christy2011,Bennacer2014,Kim2018,Edwards2018,Hack2021,Diddens2021}.  
In solutions, the composition of the solution changes in time by evaporation, 
and therefore the physical parameters characterizing the fluid, such as 
viscosity,  surface tension and the equilibrium contact angle 
change in time.  This makes the  time dependence of the 
contact angle complicated.

In a simple case, the time dependence of the contact angle can be understood
by the composition change of the droplet due to the difference in the evaporation
rate of each component. For example, in the ethanol/water mixture, the contact 
angle was observed to  increase in time~\cite{Sefiane2003,Cheng2006,Liu2008,
Diddens20171,Diddens20172,Diddens20173} and this has been understood 
as the result of the enrichment of water
which is less volatile and has larger contact angle than ethanol.

Recently, Cira et al.~\cite{Cira2015,Benusiglio2018} reported a
phenomenon that cannot be explained by such simple reasoning.
They measured the contact angle of a droplet of propylene glycol (PG)/water
on a super-wetting glass substrate, where both pure PG and pure water
droplet spreads completely.  Since the equilibrium contact angles of both components
are zero, one expects that the contact angle $\theta(t)$ will decrease monotonically
in time.  Unlike the expectation, they  observed that $\theta(t)$ stays at a  constant 
value in the first several minutes as if there is some apparent contact angle,
or even increases in time, showing maximum, and finally starts to decrease to zero.
Similar behavior has been reported by Li et al.~\cite{Li2018}
for the aqueous solution of 1,2-hexanediol.  Karpitschka et al.~\cite{Karpitschka2017}  
also reported that in water/carbon diol solution, the
contact angle indeed remains at a constant value for an extended period of time
before it starts to decrease. 

Such anomalous time dependence of the contact angle has been considered 
to be caused by the Marangoni flow: since the water content at the droplet edge
decreases quickly, the fluid flows from the edge region to the center due to the
difference in the surface tension of water and  PG.  This opposes
the spreading of the droplet and tends to increase the contact angle.

Theoretical study of such phenomena becomes complex
since the fluid flow is coupled with the concentration field through the concentration
dependence of the evaporation rate, and the  surface tension.
Diddens et al.~\cite{Diddens20171,Diddens20172,Diddens20173}
conducted a finite  element simulation for the evaporation of multi-components
droplets, accounting for the evaporation-induced Marangoni flow and thermal 
effects.  Karpitschka et al.~\cite{Karpitschka2017} studied the phenomena
for 1,2-butanediole/water and found that the apparent  contact angle $\theta_{\rm{app}}$
decreases with the increase of  the relative humidity $RH$, 
satisfying the relation $\theta_{app}\sim \left(RH_{eq}-RH\right)^{1/3}$, and 
developed a scaling  argument to explain the relation.    
Williams et al.~\cite{Williams2021} conducted more detailed analysis 
taking into account of the thermal Marangoni effect.  

In such theoretical studies,  the analysis was done by solving
a set of non-linear pde (the convective diffusion equation taking into account of the 
Marangoni effect).  Though one can do detailed modeling by
including various effects, such analysis introduces many parameters in the theory,
and makes it difficult to have overview or physical insight for the phenomena. 

In this paper, we take a different approach.  We assume that the droplet 
shape is a paraboloid, and derive time evolution equations for the parameters 
characterizing the profile, i.e., the contact radius $R(t)$ and
contact angle $\theta(t)$ using Onsagers's variational principle~\cite{Onsager19311,Onsager19312,Doi2013,Doi2015,Doi2021,Xu2015, Man2016}.  
This work is an extension of our previous work for the droplet motion induced by 
surface tension gradient~\cite{Man2017}.   In the previous problem, 
the surface tension gradient is known.  In the present problem, the
surface tension gradient is not known and must be determined by the theory.  
In this paper, we shall reconstruct the previous theory by introducing a new variable, the
concentration field of volatile component  $C(r,t)$ ($r$ being the distance from the center),
and derive a set of equations to determine their time evolution.
The theory quantifies the existing argument for the 
Marangoni contraction of evaporating droplets, and predicts how the 
initial concentration and the evaporation condition affects the time evolution of 
the contact angle. The results explains the experimental results reported previously~\cite{Cira2015, Karpitschka2017, Benusiglio2018, Li2018}.

\begin{figure}[h!]
\begin{center}
\includegraphics[bb=0 0 663 386, scale=0.5,draft=false]{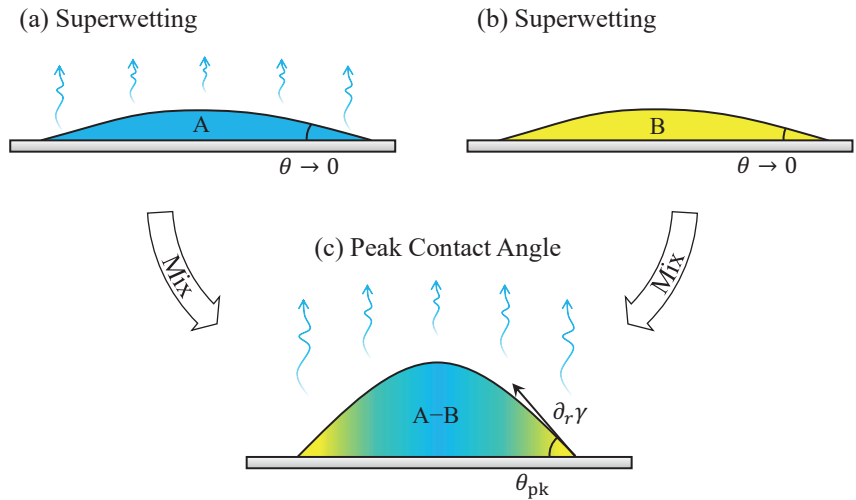}
\caption{ Schematic pictures of droplet with (a) single volatile component, (b) single nonvolatile component, and (c) mixture of the two
components. Both of the two components are complete wetting on the substrate, and the surface tension of the volatile component is larger than the nonvolatile one.  The evaporation of volatile component of the binary droplet in (c) causes an inward Marangoni flow,  which induces a peak contact angle $\theta_{\rm{pk}}$ during evaporation.}
\label{fig:drop}
\end{center}
\end{figure}

\section{Theoretical Framework} \label {sec.2}

We consider a two component droplet made of volatile component A and non-volatile
component B,  placed on a super wetting  substrate (see Figure~\ref{fig:drop}).  
We assume that the droplet contact angle is small and the surface profile is 
represented by a parabolic function, 
\begin{equation} \label{eqn:height}
  h(r,t)=H(t)\left[1-\frac{r^2}{R^2(t)}\right],
\end{equation}
where $H(t)$ is the height at the center and $R(t)$ is the radius of the droplet base. 
 The droplet volume $V(t)$ is then given by,
\begin{equation} \label{eqn:volume}
       V(t)=\frac{\pi}{2} H(t) R^2(t).
\end{equation}
The contact angle $\theta(t)$ is given by $- \partial h(r,t)/\partial r$ at $r=R(t)$.  Eqs.~(\ref{eqn:height}) and (\ref{eqn:volume}) then give,
\begin{equation}\label{eqn:angle}
     \theta(t)=\frac{4V(t)}{\pi R^3(t)}.
\end{equation}

The droplet volume changes due to evaporation.  Let $J(r,t)$ be the evaporation rate 
(defined as the liquid volume evaporating to air per unit time per unit surface area) at point $r$ 
and time $t$. Then the volume change rate $\dot V(t)$ is given by
\begin{equation} \label{eqn:vdot}
  \dot V(t)=-\int_0^{R(t)} 2\pi rJ(r,t) dr.
\end{equation}

We assume that the solution is ideal, and use the following simple model for the
evaporation rate $J(r,t)$ at point $r$~\cite{Diddens20173,Parisse1997,Eales2016}
\begin{equation}\label{eqn:j}
     J(r,t)=J_A(t) [C(r,t)- RH  ],
\end{equation}
where $C(r,t)$ is the molar fraction of the volatile component (the non-volatile
component is assumed not to evaporate),  $RH$ is the relative humidity 
and $J_A(t)$ is the  evaporation rate for the hypothetical situation of $C(r,t)=1$ and
$RH=0$, i.e., the situation that the droplet is made of pure A component and 
placed in the environment of zero humidity.

Due to the evaporation of the droplet itself, $RH$ in eq.(\ref{eqn:j}) can be a function $r$, but we ignore this effect and 
assume that $RH$ is constant independent of position and time. 
Equation (\ref{eqn:j}) indicates that evaporation takes place only when $C(r,t)> RH$.
When $C(r,t)= RH$, there is no evaporation, and when $C(r,t)<RH$, $J(r,t)$ becomes negative
and condensation takes place.

$J_A(t)$ in eq.(\ref{eqn:j}) is inversely proportional to the droplet radius $R(t)$, 
and can be written as~\cite{Parisse1997,Kobayashi2010,Man2016,Eales2016}
\begin{equation}\label{eqn:j1a}
J_A(t)= J_{A0} \frac{R_0}{R(t)} 
\end{equation}
where $J_{A0}$, and $R_0$ are the initial values of $J_A(t)$ and $R(t)$.   

We define the characteristic time of evaporation $\tau_{\rm ev}$ by
\begin{equation}\label{eqn:j1a}
    \tau_{\rm{ev}}=-\frac{V_0}{\dot V_0}
\end{equation}  Since $\dot V_0$ and $V_0$ are given by
$\dot V_0 =\pi R_0^2 J_{A0}$ and  $V_0=(\pi/4) \theta_0 R_0^3$, 
$\tau_{\rm{ev}}$ is written as $\tau_{\rm{ev}}=4 J_{A0}/\theta_0 R_0$. 
Hence Eq.(\ref{eqn:j}) is written as
\begin{equation}\label{eqn:j1}
     J(r,t)=\frac{\theta_0 R^2_0}{4\tau_{\rm ev} R(t)} [C(r,t)- RH].
\end{equation}

The conservation equation for the liquid  volume is written as,
\begin{equation} \label{massa}
   \frac{\partial h}{\partial t}  = - \frac{1}{r} \frac{\partial (rv_f h)}{\partial r}  - J,
\end{equation}
where $v_f (r)$ is the height averaged velocity of the fluid at point $r$. 

By use of eq.( \ref{eqn:height}), the left hand side of Eq.~(\ref{massa}) can be 
expressed by $\dot R$ and $\dot V$. Then  Eq.~(\ref{massa})  is solved for $v_f(r)$
(see Supplemental Material for details).  This gives
\begin{equation} \label{vf}
  v_f=r\frac{\dot R}{R}
           -\left(\frac{r}{2V}+\frac{r^3}{2\pi R^4h}\right)\dot V-\frac{1}{rh}
                            \int_0^r r'J(r';t)dr'.
\end{equation}
 The mass conservation equation for the volatile component is written as
\begin{equation} \label{massb}
   \frac{\partial{\left(Ch\right)}}{\partial t} = - \frac{1}{r} \frac{\partial (rv_f Ch)}{\partial r}-J,
 \end{equation}
The first term on the right hand side represents the effect of convection, and the second term represents the evaporation. The effect of diffusion is ignored in the present theory, but its effect will be discussed later.

Combining Eqs.~(\ref{massa}) and (\ref{massb}), we have the time derivative of $C(r,t)$,
\begin{equation}   \label{massc}
   \dot C = - v_f \frac{\partial C}{\partial r} - \frac{J}{h}\left(1-C\right).
\end{equation}
This equation indicates explicitly that even if the initial composition of the droplet is uniform, the
composition becomes non-uniform due to the evaporation of volatile component. 
Since $h(r,t)$ becomes zero at the contact line, Eq.(\ref{massc}) also indicates that 
at  $r=R(t)$, $C(r,t)$ relaxes to the equilibrium value very quickly, and therefore 
$C(R(t),t)=RH$. 

Careful inspection of the above set of equations indicates that there is only 
one unknown,  $\dot R$, which we need to calculate to determine the time 
evolution of the system.  This is seen as follows. 
Suppose that we know $V(t)$,  $R(t)$ and $C(r,t)$ at time $t$, then 
$J(r,t)$ is given by Eq.~(\ref{eqn:j1}) and $\dot V$ is given by Eq.~(\ref{eqn:vdot}). Hence 
 $v_f(r)$ is expressed as a linear function of $\dot R$ by Eq.~(\ref{vf}).  
Therefore if $\dot R$ is known,  $\dot C$ is calculated by Eqs.~(\ref{vf}) and  
(\ref{massb}). Therefore, the time evolution of $R(t)$, $V(t)$ and $C(r,t)$ 
can be calculated if $\dot R$ is known.

In the following, we shall use the Onsager variational principle  
\cite{Onsager19311,Onsager19312,Doi2013,Doi2015, Doi2021,Xu2015, Man2016,Man2017,Wu2018,Wu2019,Jiang2020,Yang2021} to derive the equation for  $\dot R(t)$.  The calculation is 
essentially the same as that used in the previous paper \cite{Man2017} 
for the motion of an evaporating droplet under surface tension gradient.   
We consider the Rayleihian $\Ray$ defined by
\begin{equation}\label{ray}
      \Ray =\Phi+\dot F,
\end{equation}
where $\Phi$ is the energy dissipation function, and $\dot F$ is the free energy 
time change rate.  We obtain $\Ray$ as a function of $\dot R$, and obtain
$\dot R$ by the condition $\partial \Ray/\partial \dot R =0$.  

The energy dissipation is caused by the fluid flow inside the droplet 
and $\Phi$ is written as \cite{Man2017},  
\begin{equation}\label{phi}
      \Phi =\int_0^R 2\pi r \frac{\eta}{2h}\left[12\left(v_f-\frac{v_s}{2}\right)^2+v_s^2\right]dr,
\end{equation}
where $\eta$ is the viscosity of the fluid, which is assumed  
 to be constant, and $v_s(r)$ is the fluid 
velocity at the liquid-vapor interface.  In Eq.~(\ref{phi}), we have 
ignored the extra friction associated with the contact line motion (i.e., the
contact line friction coefficient  $\xi_{\rm{cl}}$
\cite{ Man2016} has been set to zero.)

The interfacial free energy of the system is given by
\begin{equation}  \label{free_energu20}
   F = \int_0^R 2 \pi r dr \left[ \gamma(r) \left( 1 + \frac{1}{2} h'^2 \right ) -  \gamma_{A}\cos\theta_{eA}  \right],
\end{equation}
where $\theta_{eA}$  is the equilibrium contact angle for A component, and 
$\gamma(r)$ is the surface tension of the solution at point $r$.  
We assume a linear 
dependence  of  $\gamma(r)$ on $C(r)$~\cite{Eales2016},
\begin{equation} \label{eqn:surface_tension1}
       \gamma(r)=C(r) \gamma_A+  \left[1- C(r)\right] \gamma_B,
\end{equation}
where $\gamma_A$ and $\gamma_B$ are the  surface tensions of pure $A$ and $B$ components. 

The dissipation function $\Phi$ involves the surface velocity $v_s(r)$,  but 
this is expressed by $v_f(r)$ and the surface tension gradient
$\partial \gamma/\partial r$ (see Supplemental Material).
\begin{equation} \label{vs}
v_s=\frac{3v_f}{2}+\frac{h}{4\eta}\frac{\partial \gamma}{\partial r}.
\end{equation}

By use of Eqs.~(\ref{vf}) and (\ref{vs}) in Eq.~(\ref{phi}), the Rayleighian can be
expressed as a quadratic function of $\dot R$. Therefore the time evolution for
$R(t)$ is determined by the condition $\partial \Ray/\partial \dot R=0$. 
The calculation is straightforward, but cumbersome, and described in the Supplemental Material.  
In the end, we obtain
the following equation for $\dot R$
\begin{eqnarray}\label{eqn:rdot}
\begin{aligned}
  \dot R=&\frac{\left(\gamma_{\rm{re}}-1\right)\theta V_0^\frac{1}{3}}{3\gamma_{\rm{re}}\alpha \tau_{\rm{re}}}
\left[\int_0^R C\frac{r}{R^2}
\left(\frac{2r^2\theta^2}{R^2}-5\right)dr+\frac{\theta^2-\gamma_{\rm{re}}\theta_{\rm eA}^2}{2\left(\gamma_{\rm{re}}-1\right)}+
\frac{3}{2}C\left(R\right)+1\right]
-\frac{R\dot V}{4\alpha V}\\-&\frac{\theta \theta_0R_0^2}{4\alpha R^3 \tau_{\rm{ev}}}
\int_0^R \frac{r}{h^2}\left[\int_r^Rr'\left(C- RH \right)dr'\right]dr,
\end{aligned}
\end{eqnarray}
where $\gamma_{\rm{re}}=\gamma_A/\gamma_B$ is the surface tension ratio of the two components, $\alpha = \ln (R/2 \epsilon) -1$ is a parameter which is regarded as constant in the subsequent analysis, $\epsilon$ is the molecular cutoff length, and $\tau_{\rm{re}}$ is defined by
\begin{equation}
                 \tau_{\rm{re}} = \eta V_0^{1/3}/\gamma_A
\end{equation}
which represents the characteristic relaxation time of the droplet determined by viscosity $\eta$
and the surface tension $\gamma_A$.

Since $\dot\theta$ is related to $\dot V$ by (see Eq.~(\ref{eqn:angle})),
\begin{equation}\label{eqn:dottheta}
\begin{aligned}
\dot \theta&=\frac{\theta\dot V}{V}-\frac{3\theta\dot R}{R}.
\end{aligned}
\end{equation}
Eq.~(\ref{eqn:rdot}) gives the following  equation for $\dot \theta$,
\begin{eqnarray}\label{eqn:thetadot}
\begin{aligned}
\dot \theta
            =&\frac{\left(\gamma_{\rm{re}}-1\right)\theta^2 V_0^\frac{1}{3}}{\gamma_{\rm{re}}\alpha \tau_{\rm{re}}R}
\left[\int_0^R C\frac{r}{R^2}\left(5-\frac{2r^2\theta^2}{R^2}\right)dr
-\frac{\theta^2-\gamma_{\rm{re}}\theta_{\rm eA}^2}{2\left(\gamma_{\rm{re}}-1\right)}
-\frac{3}{2}C\left(R\right)-1\right]\\
+&\frac{\theta\dot V}{V}\left(1+\frac{3}{4\alpha }\right)
+\frac{3\theta^2 \theta_0R_0^2}{4\alpha R^4\tau_{\rm ev}}\int_0^R \frac{r}{h^2}\left[\int_r^Rr'\left(C-RH\right)dr'\right]dr.
\end{aligned}
\end{eqnarray}

To summarize,  the time evolution 
of the system can be calculated as follows.  The non-equilibrium state of the system is
specified by three variables $\theta(t)$ and $V(t)$ and $C(r,t)$. 
Their time derivatives are calculated as follows.
\begin{itemize}
\item[(1)] $\dot V$ is  calculated by Eq.~(\ref{eqn:vdot}), where $J(r,t)$ is given by Eq.~(\ref{eqn:j}).
\item[(2)] $\dot \theta$ and $\dot R$ are given by Eq.~(\ref{eqn:thetadot}) and Eq.~(\ref{eqn:dottheta})
\item[(3)]  $\dot C$ is given by Eq.~(\ref{massc}), where $v_f$ is given by Eq.~(\ref{vf})
\end{itemize}
Since they are a set of ordinary and first order partial differential equations, 
the calculation is simple and quick. We shall show the results 
in the next section.

The above time evolution equations become simple 
for the special case of pure liquid (i.e., the case of $C=1$). 
In this case,  taking $\theta_{\rm eA}=0$ they reduce to the following equations (see Supplemental Material),
\begin{eqnarray}
  \dot R &=& \frac{V_0^{1/3}}{6 \alpha \tau_{re}} \theta^3 + \frac{R}{4V}\dot V 
                                                             \label{pure1}
                                \\
  \dot \theta &=& -\frac{V_0^{1/3}}{2 \alpha \tau_{re}} \theta^4 + \frac{\theta }{4V}\dot V
                                                              \label{pure2} 
\end{eqnarray}
The first terms on the right hand side of Eqs.~(\ref{pure1}) and (\ref{pure2}) represent
the Tanner's law~\cite{Tanner1979,Oron1997} for spreading, and the second terms represents the effect of liquid 
evaporation.  These equations agree with the previous results~\cite{Man2016}.  
Equation~(\ref{pure2}) indicates that $\dot \theta$ is always negative for pure
liquid droplet.  This conclusion is not consistent with the recent works~\cite{Jambon-Puillet2018,Eggers2010}
which claim that $\dot \theta$ become zero in a certain range of time
if the droplet is evaporating.  The discrepancy can be resolved by taking into 
account of the non-uniform evaporation rate in pure liquid droplet. Detailed 
discussion will be given in a future article.

\section{Results and Discussions} \label {sec.3}

Since there are many parameters involved in Eq.~(\ref{eqn:thetadot}), we focus
on two parameters which can be changed easily in experiments, the initial
composition of the volatile component $C_0$, and the humidity $RH$.  
In the following we study the time evolution of $\theta(t)$ and $R(t)$ by changing
the two parameters while keeping the other parameters fixed as follows;  
initial contact angle $\theta_0=0.3$,  
surface tension ratio $\gamma_{\rm{re}}=\gamma_A/\gamma_B=2$, 
equilibrium contact angles $\theta_{\rm eA}=\theta_{\rm eB}=0$. 
With such set up, the subsequent discussion is limited to the case 
that the surface tension of volatile component is larger than the non-volatile 
component (i.e., the case of $ \gamma_{\rm{re}} >1 $), but the opposite case 
of $\gamma_{\rm re}<1$ can be studied by the present theory and is discussed
in Supplemental Material.

\subsection{Effect of initial composition $C_0$}

\begin{figure}[h!]
\begin{center}
\includegraphics[bb=0 0 501 264, scale=0.92,draft=false]{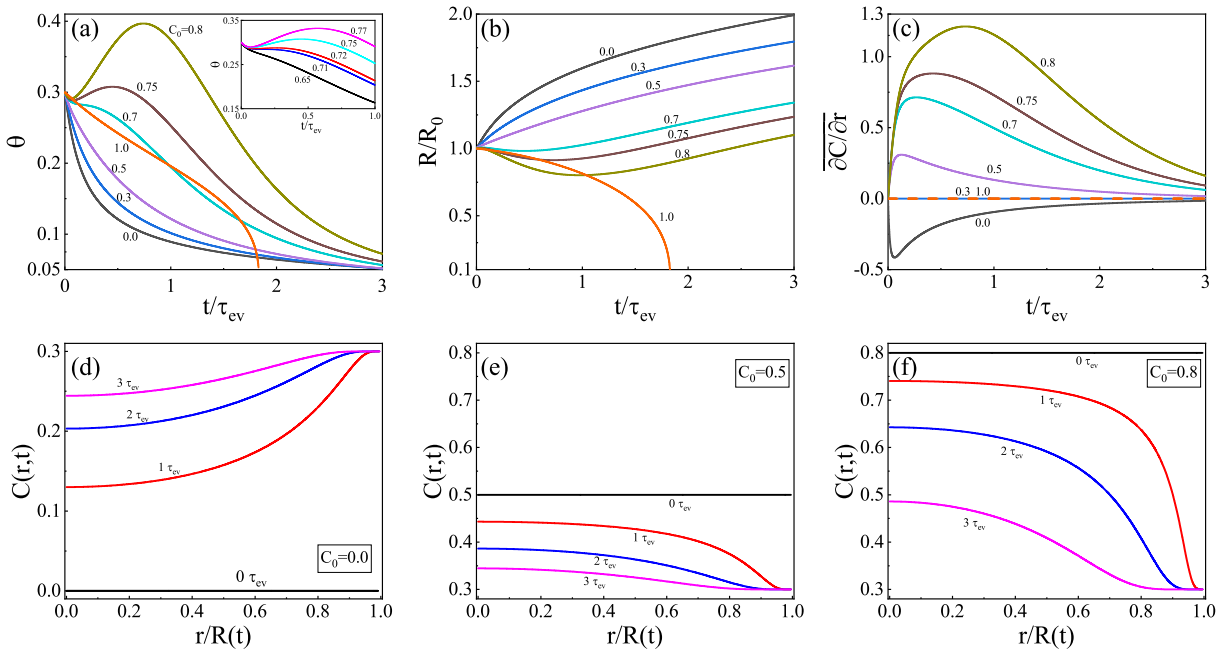}
\caption{Evolution of (a) droplet contact angle $\theta(t)$,
(b) droplet contact radius $R(t)/R_0$, and (c)  average concentration gradient $\overline{\partial C/\partial r}$ of  an evaporating binary droplet for various
initial concentration of volatile component, $C_0$. The inset figure in (a) is the zoom in evolution of the contact angle, indicating that the peak contact angle changes to quasi-stationary (or apparent) contact angle for a short time interval by decreasing $C_0$ ($C_0=0.72$ and $0.71$). (d), (e), and (f) show the concentration distribution in the droplet at different times 
for three initial concentration (d) $C_0=0.0$, (e) $C_0=0.5$, and (f) $C_0=0.8$. In all  cases,  $k_{\rm{ev}}=0.002$, and $RH=0.3$.}
\label{fig:evolution}
\end{center}
\end{figure}

Figure~\ref{fig:evolution} shows the time variation of the contact angle 
$\theta(t)$ and the contact radius $R(t)$  for various initial concentration $C_0$
of volatile component.  

The curves of  $C_0=0.3 $ correspond to the
case of non-evaporating droplet since the flux $J$ in Eq.~(\ref{eqn:j}) becomes zero 
at this concentration.  In this case, the droplet keeps uniform composition and behaves
as a non-volatile liquid.  Its dynamics is described by Eqs.~(\ref{pure1}) and (\ref{pure2}) 
with $\dot V=0$. Hence, $R(t)$ increases in time and $\theta(t)$ decreases in time.

The curves of  $C_0=1.0 $ correspond to the case of pure evaporating droplet
and its dynamics is again described by Eqs.~(\ref{pure1}) and (\ref{pure2}).
In the case of Figure~\ref{fig:evolution} (a) and (b), both  $\theta(t)$ and $R(t)$ 
decrease to zero monotonically.  The monotonic decrease of $\theta(t)$ 
is a general result of the  present theory for pure liquid, but 
the monotonic decrease of $R(t)$ is a result of the choice 
of the initial condition.  (The figure shows the result of the case of 
$\theta_0=0.3$.  If $\theta_0$ is larger, $R(t)$ first increases in time and then 
starts to decrease.)

The droplets having initial concentration $C_0$ between 0.3 and 1, 
show the effect of Marangoni
contraction.  As it is seen in Figure~\ref{fig:evolution} (e) and (f), 
$C(r,t)$  decreases quickly to the equilibrium 
value $0.3$ at the edge of the droplet, while $C(r,t)$ at the center is larger than this
value.  This concentration gradient creates surface tension gradient in the droplet, and
causes an inward Marangoni flow from the edge to the center.  Such inward flow 
suppresses the decrease of  $\theta(t)$, and creates the plateau-like region in
the plot of $\theta(t)$.  The inset of Figure~\ref{fig:evolution} 
(a) shows such plateau region  for $C_0$, $0.7 < C_0 < 0.72$.  With further increase of
$C_0$,  $\theta(t)$ shows a peak.   

Although the present theory accounts for the effect of Marangoni contraction, 
it does not give a clear plateau behaviour for $\theta(t)$: the plateau 
appears only in a limited time range or limited parameter space. It is difficult to 
identify the ``apparent contact angle" in our result.  We therefore focused on the
peak of the contact angle since it represents the anomalous time dependence
of $\theta(t)$, and corresponds to the height of the plateau 
when $\theta(t)$ shows the plateau. Cira also reported that the quasi-stationary
contact angle is observed in a limited parameter space~\cite{Cira2015}.

The anomaly in the contact angle is caused by the concentration 
gradient created in the binary droplet by evaporation.  The relation between the
concentration gradient and the peak contact angle  is seen 
in Figure~\ref{fig:evolution} (c), where the average
concentration gradient
\begin{equation}
    \overline{\frac{\partial C}{\partial r}} = \frac{2}{R^2} \int_0^R dr r \frac{\partial C}{\partial r} 
\end{equation}
is plotted against time. It is seen that  $\overline{\partial C/\partial r}$ has a peak, and that
the peak time of $\overline{\partial C/\partial r}$ is close to the peak time of $\theta(t)$.  
This again supports the mechanism of the contact angle anomaly in binary droplet.

If the initial concentration $C_0$ is smaller than the equilibrium value 0.3,  condensation
takes place instead of evaporation.  It is interesting to see that there is no anomaly in  the
behavior of $\theta(t)$ and $R(t)$: their behaviour are quite continuous across this boundary 
from evaporation to condensation.  

\subsection{Effects of humidity}

\begin{figure}[h!]
\begin{center}
\includegraphics[bb=0 0 499 280, scale=0.92, draft=false]{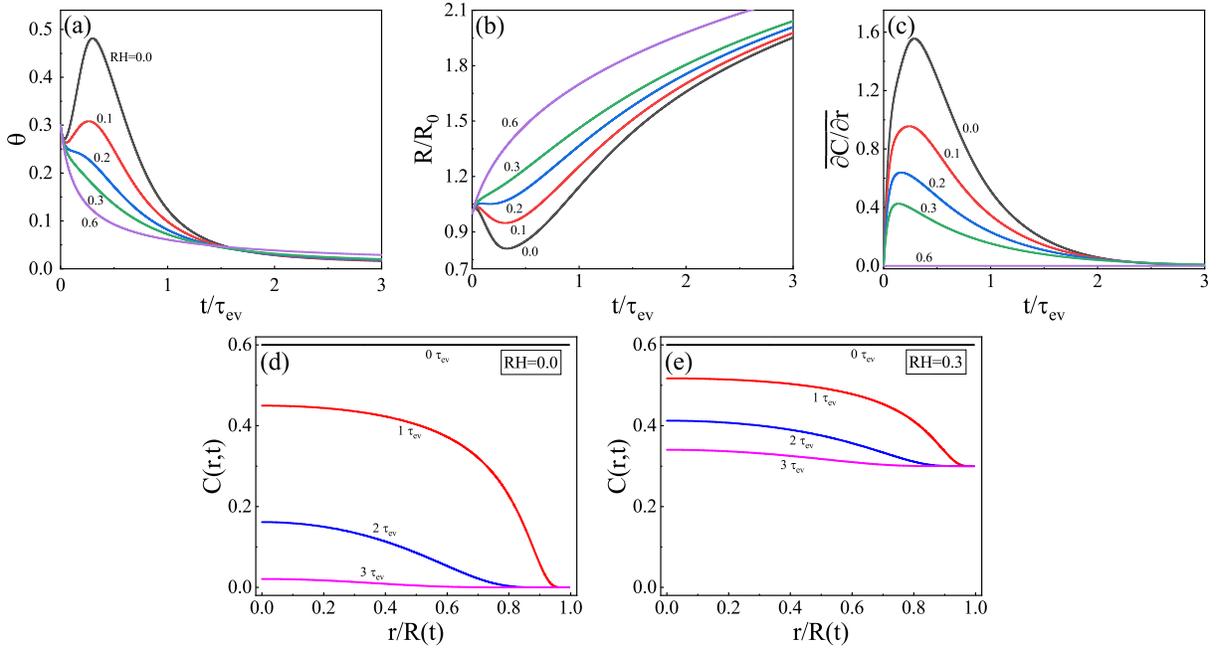}
\caption{ (a) Time dependence of (a) the contact angle $\theta(t)$, (b) the contact radius
$R(t)/R_0$, and (c) the averaged concentration gradient of the volatile component for various
relative humidity $RH$. The evolution of the distribution of volatile component concentration within the droplet for (d) $RH=0.0$, and (e) $RH=0.3$. In all cases, $C_0=0.6$, and $k_{\rm ev}=5\times10^{-4}$.}
\label{fig-rh}
\end{center}
\end{figure}

Figure~\ref{fig-rh} shows the effect of humidity $RH$ on the time evolution of the 
droplet having initial composition $C_0=0.6$.  The effect of humidity $RH$ is qualitatively similar to the effect of initial composition  $C_0$.  When humidity is high ($RH=0.6$), there is no evaporation, and the droplet  spreads 
on the surface as a non-volatile liquid, showing monotonous decrease of 
$\theta(t)$ and monotonous increase of $R(t)$.  With the decrease of $RH$, 
evaporation starts, and the effect of Marangoni contraction sets in.
Accordingly $\theta(t)$ starts to show a plateau or a peak.  The time evolution 
of the average concentration gradient (Figure~\ref{fig-rh}(c))  and the time 
evolution of the concentration profile (Figure~\ref{fig-rh} (d) and (e) ) confirms
that the same mechanism is working as in the case of  Figure~\ref{fig:evolution}.

\subsection{Peak contact angle}

\begin{figure}
\begin{center}
\includegraphics[bb=0 0 335 133, scale=1.2, draft=false]{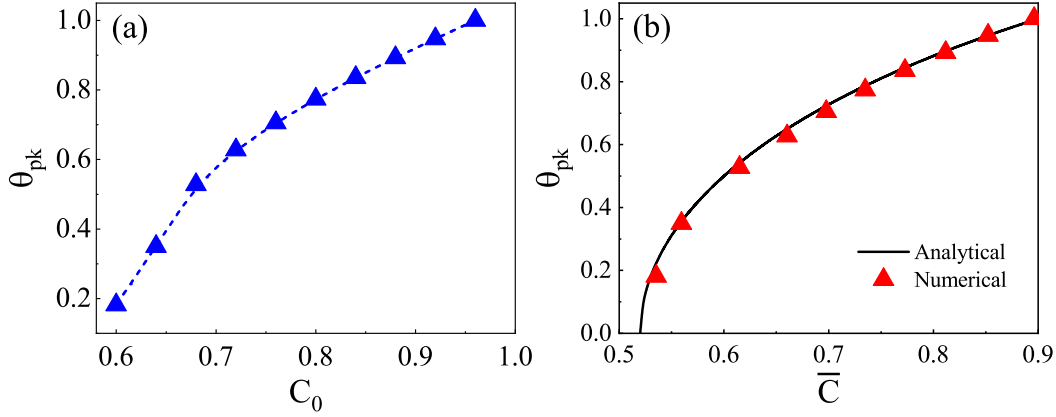}
\caption{  (a) The dependence of the peak contact angle $\theta_{\rm{pk}}$ on the initial concentration of volatile A-component, $C_0$.  (b) The dependence of $\theta_{\rm {pk}}$ on $\bar C$, where $\bar C$ is the average concentration at the moment when $\theta_{\rm{pk}}$ occurs, $\bar C=2 \int_0^R Cr/R^2 dr$. The results obtained from the full numerical calculations are denoted by symbols, while the analytical result is denoted by the black-solid line. All the other parameters are $k_{\rm ev}=1\times10^{-4}$, and $RH=0.2$.
. }
\label{fig:thetaappa}
\end{center}
\end{figure}

As it has been shown in previous sections, the contact angle $\theta(t)$ shows a peak $\theta_{\rm{pk}}$ when the evaporation is strong (i.e., when the initial concentration $C_0$ is large or when the humidity $RH$ is small).  Figure~\ref{fig:thetaappa} (a) 
shows the plot of the peak contact angle $\theta_{\rm{pk}}$  against the
initial concentration $C_0$.   Figure~\ref{fig:thetaappa} (b) shows the plot
of $\theta_{\rm{pk}}$ against  the average
concentration $\bar C=2 \int_0^R Cr/R^2 dr$ when the peak appears. 

In order to understand the behavior shown in Figure~\ref{fig:thetaappa}  (b), 
we evaluate the integral on the right hand side of Eq.~(\ref{eqn:thetadot}) 
by replacing $C(r)$ with the average value $\bar C$. We also 
assume that $C(R)=RH$.  This approximation
gives the following equation for $\dot \theta$.
\begin{equation}\label{eqn:thetadot1}
    \begin{aligned}
\tau_{\rm{ev}}\dot \theta=\frac{\left(\gamma_{\rm{re}}-1\right)\theta^2 V_0^\frac{1}{3}}{\gamma_{\rm{re}}\alpha k_{\rm{ev}}R}
\left[ \bar C\left(\frac{5}{2}-\frac{\theta^2}{2}\right)
-\frac{\theta^2-\gamma_{\rm{re}}\theta_{eA}^2}{2\left(\gamma_{\rm{re}}-1\right)}-\frac{3}{2}RH-1\right]
+\frac{\theta\dot V\tau_{\rm{ev}}}{4V}
\end{aligned}
\end{equation}
At the peak position, $\dot \theta$ is equal to zero. To simplify the equation further, 
we consider the limit of $k_{\rm{ev}}\rightarrow 0$. Then Eq.~(\ref{eqn:thetadot1})
is simplified as
\begin{equation}\label{eqn:thetaappa1}
    \begin{aligned}
 \bar C\left(\frac{5}{2}-\frac{\theta_{\rm{pk}}^2}{2}\right)
-\frac{\theta_{\rm{pk}}^2-\gamma_{\rm{re}}\theta_{eA}^2}{2\left(\gamma_{\rm{re}}-1\right)}-\frac{3}{2}RH-1=0.
\end{aligned}
\end{equation}
Equation (\ref{eqn:thetaappa1}) is solved for $\theta_{\rm{pk}}$ as
\begin{eqnarray}
\theta_{\rm{pk}} &=&\left[  \frac{ 
                                   (\gamma_{\rm{re}}-1) ( 5\bar C-3RH -2)
                                      + \gamma_{\rm{re}}\theta_{eA}^2
                              }
                              {1 +\left(\gamma_{\rm{re}}-1\right)\bar C}
                \right]^\frac{1}{2}
                                 \label{eqn:thetaappa2}     \\
                  &=& \left[
                           \frac{(\gamma_A-\gamma_B)(5\bar C_A - 3RH -2)
                                            +\gamma_{A}\theta_{eA}^2 }
                                 {\gamma_A\bar C_A+\gamma_B\bar C_B}
                       \right] ^\frac{1}{2}
                                          \label{eqn:thetaappa3}.
\end{eqnarray}
where $\bar C_A$ and $\bar C_B$ stand for $\bar C$ and $1-\bar C$.

We compare the value of $\theta_{\rm{pk}}$ calculated by Eq.~(\ref{eqn:thetaappa3}) 
with that obtained by the full numerical calculation in Figure~\ref{fig:thetaappa}(b) 
for $RH=0.2$.  It is seen that the result of Eq.~(\ref{eqn:thetaappa3}) (solid line) agrees
quite well with that of the full numerical calculation (data points). 

Equation~(\ref{eqn:thetaappa3}) indicates that for the peak (or plateau)  
to be observed, $5\bar C - 3RH -2$ must be positive. Since $\bar C$ is smaller 
than $C_0$, this condition indicates that 
$5C_0 - 3RH -2$ must be positive for the peak to be observed. In other
word, for the peak (or plateau) to be observed the initial condition must satisfy
\begin{equation}
     C_0 > \frac{3 RH +2}{5}
\end{equation}
Such condition is consistent with the behaviour of $\theta(t)$ shown in  
Figure \ref{fig:evolution}.  It is interesting to check the  prediction by experiments.

Equation (\ref{eqn:thetaappa3}) is useful to estimate the value of $\theta_{\rm{pk}}$.
Since $\bar C$ at the peak position is not very different from the initial concentration $C_0$,
$\theta_{\rm{pk}}$ can be estimated by Eq.~(\ref{eqn:thetaappa3}) with
$\bar C$ replaced by $C_0$. 

\begin{figure}[h!]
\begin{center}
\includegraphics[bb=0 0 333 268, scale=0.7, draft=false]{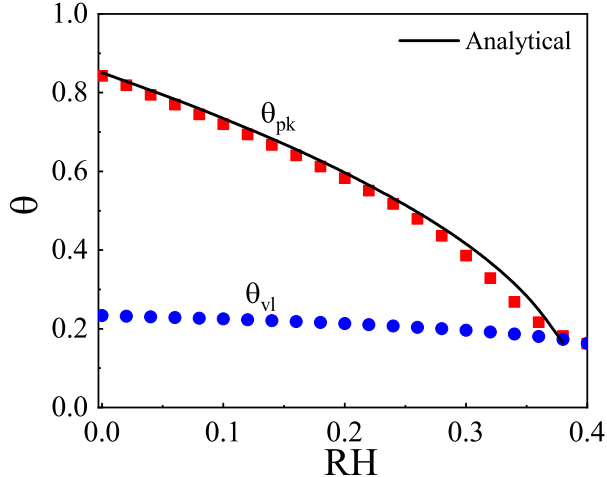}
\caption{
The dependence of the peak contact angle $\theta_{\rm{pk}}$ and the valley contact angle $\theta_{\rm{vl}}$ on the relative humidity of volatile A-component, $RH$. Symbols denote the results of the full numerical calculations. The black-solid line denotes the analytical results of Eq. (\ref{eqn:thetaappa2}), where we take $\bar C=0.637$ and $\theta_{\rm eA}=0$. Here the value of $\bar C$ is the average of $\bar C$ when $\theta_{pk}$ occurs for different $RH$.  In all cases, $k_{ev}=1\times10^{-4}$, and $C_0=0.7$. }
\label{fig:rh-theta}
\end{center}
\end{figure}

Equation~(\ref{eqn:thetaappa3}) also indicates that $\theta_{\rm pk}$ depends on $RH$ in
the form of $\theta_{\rm pk} \simeq (RH_{eq} - RH )^{1/2}$. This relation has been 
confirmed by our full numerical calculations as it is shown in Figure~\ref{fig:rh-theta}. 
The relation is also qualitatively consistent with the previous experimental results. Karpitschka et al.~\cite{Karpitschka2017} reported that the apparent contact angle $\theta_{app} $ decreases with $RH$ following the relation $(RH_{eq} – RH )^{1/3}$.  Cira et al.~\cite{Cira2015} found a linear relation between $\cos \theta_{app} $ and $RH$, which is equivalent to $\theta_{app} \propto (RH_{eq} – RH )^{1/2}$. When $\theta_{\rm pk}$ meets the valley contact angle $\theta_{\rm vl}$, a plateau of the contact angle appears. 

\section{Conclusion}
In this paper, we developed a theory for the spreading dynamics of an evaporating droplet 
made of volatile and nonvolatile components. We have shown that in such two component
liquid, the evaporation makes the time variation of the contact angle $\theta(t)$ 
and contact radius $R(t)$ quite complex since evaporation of the volatile component
makes the concentration in the droplet non-uniform, and creates a Marangoni flow in the
droplet.  Here we limited our discussion to the case that the surface tension of the
volatile component is larger than that of non-volatile component.  We also limited
the discussion that the equilibrium contact angles of both components are zero (super
wetting surface).  In such a case, the induced Marangoni flow is directed inward (from the edge 
to the center), which tends to decrease the contact radius and to increase the contact angle.  

Using the Onsager principle, we derived a set of equations 
which determine the time evolution of the contact angle $\theta(t)$, 
contact radius $R(t)$ and concentration of volatile component $C(r,t)$.
To derive the equation, we assumed
\begin{itemize}
\item[(1)] The droplet has a profile of parabola, characterized by $R(t)$ and $\theta(t)$.
\item[(2)] The liquid is an ideal solution of A/B mixture, and the evaporation rate is given by
Eq.~(\ref{eqn:j})
\end{itemize}
Assumption (1) is needed to describe the spreading dynamics by two parameters $\theta(t)$
and $R(t)$, and assumption (2) is needed to introduce the effect of evaporation. Although the parabola assumption of the droplet surface profile can reveal a few features of drying droplets, it should be noticed that complicated shape of surface profile appears of drying droplets. Then, the full hydrodynamic equations are needed for the study of such cases. 

There are other assumptions and approximations in the theory, such as the 
ignorance of diffusion and thermal effect. Diffusion decreases the concentration gradient and weaken the Marangoni flow.  The effect of diffusion can be estimated by the P\'eclet number $Pe=v_f R/D$, where $v_f$ is the fluid velocity, $R$ is the contact radius, and $D$ is the mutual diffusion constant of water and oil. Taking the value of $v_f=2.1\times10^{-6}~{\rm m/s}$, $R= 2.5\times 10^{-3}~{\rm m}$ and $D=5.0\times 10^{-10}~{\rm m^2/s}$, $Pe$ is around 10.  Therefore the diffusion is not entirely negligible, but it will not affect the present results seriously. 

We emphasize that despite such approximations,  the present theory explains the
experimental observation that the plateau (or peak) contact angle appears only when the composition of the volatile component in the droplet is large enough, and that the value decreases with the increase of room humidity approximately in proportional to $(RH_{eq}-RH)^{1/2}$~\cite{Cira2015}. Moreover, for the cases of $\gamma_{\rm re} < 1$ that the surface tension of the faster evaporation component being smaller than the slower one~\cite{Williams2021}, the enhanced droplet spreading by increasing the initial concentration of volatile component are captured by our model.

\bigskip
{\bf Acknowledgement.}~~
We thank N. J. Cira for useful discussions. This work was supported by the National Natural Science Foundation of China (Grant No. 21822302), the joint NSFC-ISF Research Program, China (Grant No. 21961142020), and the Fundamental Research Funds for the Central Universities, China.




\end{document}


\title{Supplemental Material for \\ Contact Angle of An Evaporating Droplet of Binary Solution on a
Super Wetting Surface}
\author{Mengmeng Wu}
\affiliation{Center of Soft Matter Physics and its Applications, Beihang University, Beijing 100191, China}
\affiliation{School of Physics, Beihang University, Beijing 100191, China}
\author{Masao Doi}
\affiliation{Center of Soft Matter Physics and its Applications, Beihang University, Beijing 100191, China}
\affiliation{School of Physics, Beihang University, Beijing 100191, China}
\author{Xingkun Man}
\email{manxk@buaa.edu.cn}
\affiliation{Center of Soft Matter Physics and its Applications, Beihang University, Beijing 100191, China}
\affiliation{School of Physics, Beihang University, Beijing 100191, China}

\maketitle

\subsection{Detailed derivation of the model}

\begin{figure}[h!]
\begin{center}
\includegraphics[bb=0 0 663 386, scale=0.6,draft=false]{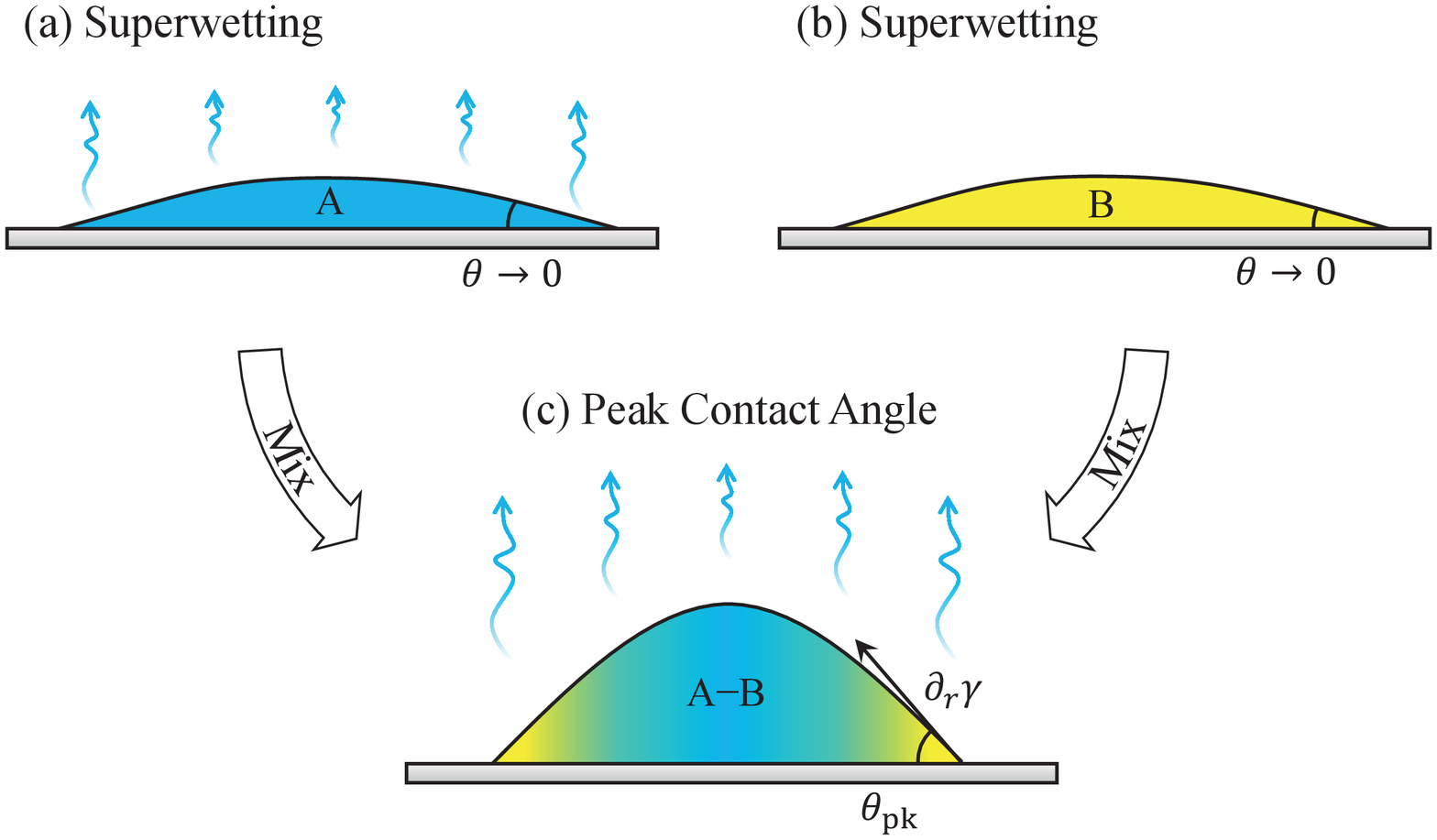}
\caption{ Schematic pictures of droplet with (a) single volatile component, (b) single nonvolatile component, and (c) mixture of the two
components. Both of the two components are super-wetting on the substrate, and the surface tension of the volatile component is larger than the nonvolatile one.  The nonuniform evaporation of the binary droplet in (c) causes an inward surface tension gradient,  which induces a peak contact angle $\theta_{\rm{pk}}$ during evaporation.}
\label{fig:drop1}
\end{center}
\end{figure}
Consider a droplet placed on a substrate which is a solution made of volatile component A and non-volatile component B, as shown in Fig.~{\ref{fig:drop1}}. We assume that the droplet contact angle is small and that the surface profile is approximated by a parabolic function
\begin{equation} \label{eqn:height}
  h(r,t)=H(t)\left[1-\frac{r^2}{R^2(t)}\right],
\end{equation}
where $H(t)$ is the height at the droplet center and $R(t)$ is the radius of the droplet base.
Then, the droplet volume $V(t)$ is given by
\begin{equation} \label{eqn:volume}
       V(t)=\frac{\pi}{2} H(t) R^2(t).
\end{equation}
The contact angle $\theta(t)$ is given by $- \partial h(r,t)/\partial r$ at $r=R(t)$.  Eqs.~(\ref{eqn:height}) and (\ref{eqn:volume}) then give
\begin{equation}\label{eqn:angle}
     \theta(t)=\frac{4V(t)}{\pi R^3(t)}.
\end{equation}
%
\begin{figure}[h!]
\begin{center}
\includegraphics[bb=0 0 517 209, scale=0.55,draft=false]{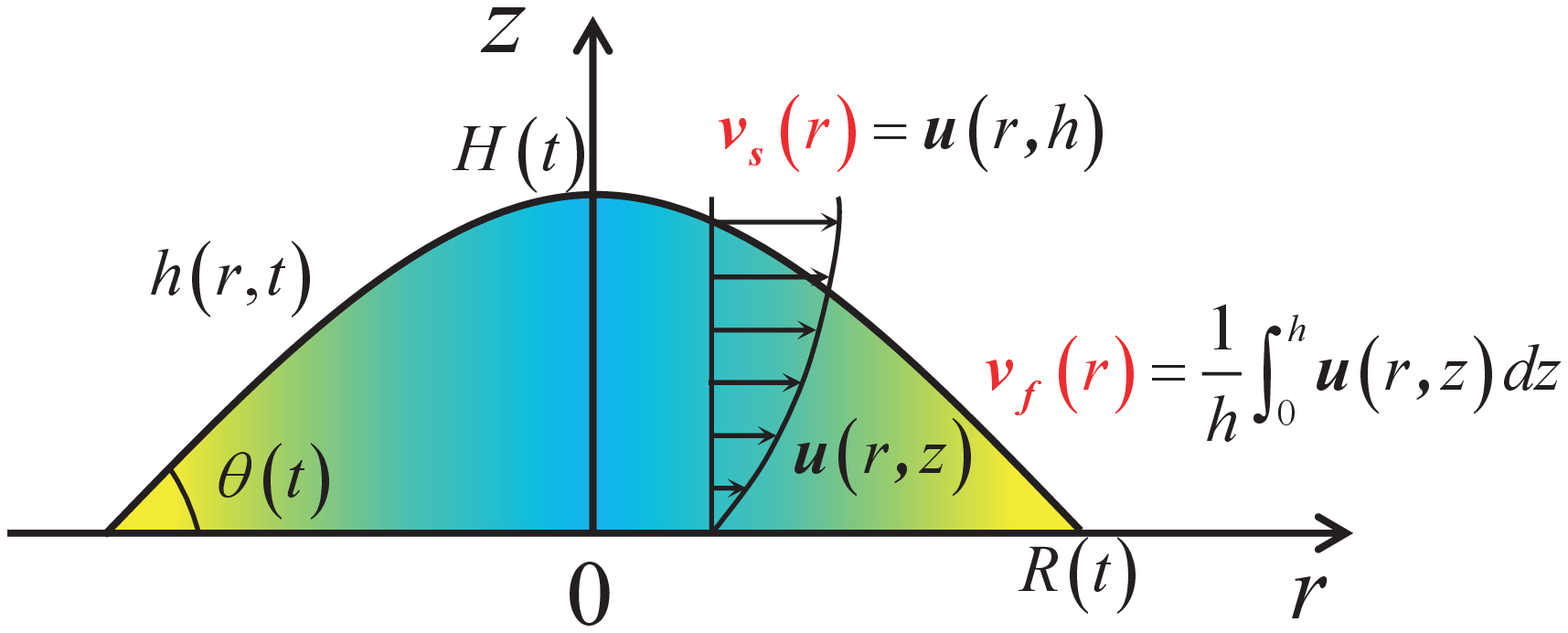}
\caption{ Schematic of the height averaged fluid velocity $v_f(r)$ and the fluid velocity at the liquid/vapor interface $v_s(r)$ (side view). The fluid flow velocity is $u(r, z)$,  the radius of the contact line is $R$, the height of the droplet at the
center is $H$, the contact angle is $\theta$, and the profile of the droplet
liquid-vapor interface is $h(r)$.}
\label{fig:drop2}
\end{center}
\end{figure}
Let $J(r)$ be the evaporation rate (the liquid volume evaporating to air per unit time
per unit surface area) at point $r$. The volume change rate is related to $J$ as
\begin{equation} \label{eqn:vdot}
  \dot V(t)=-\int_0^R 2\pi rJ dr.
\end{equation}
We assume the evaporation rate of the nonvolatile component B equals to zero, and the evaporation rate of the
binary droplet on a substrate is proportional to the concentration of the volatile component A, which is given by the following form~\cite{Parisse1997,Eales2016,Diddens20173}
\begin{equation}\label{eqn:j}
     J(r)=J_A\left[C(r)-RH\right],
\end{equation}
where $C(r)$ is the height-averaged mass fraction of the volatile component in the droplet, $RH$ is the relative humidity of A component, and $J_A$ is the evaporation rate of pure droplet having the same volume and base radius of the solution droplet when $RH=0$. Here we reduced our problems to the ideal case of Raoult's law. The explicit form of $J_A$ is written by~\cite{Parisse1997,Kobayashi2010,Man2016}
\begin{equation}\label{eqn:ja}
     J_A=\frac{\theta_0 R_0^2}{4R\tau_{\rm{ev}}},
\end{equation}
where  $\tau_{\rm{ev}}=-V_0/\dot V_0$ is the characteristic evaporation time for the pure A droplet when $RH=0$ with $\dot V_0$ being a given constant, and $V_0$, $\theta_0$ and $R_0$ are the initial values of $V(t)$, $\theta(t)$ and $R(t)$ , respectively.
The liquid/vapor surface tension $\gamma$ of the solution depends on the composition in the solution.
We assume a linear dependence on $C(r)$~\cite{Eales2016},
\begin{equation} \label{eqn:surface_tension2}
       \gamma(r) =  \gamma_A C + \frac{\gamma_A}{\gamma_{\rm{re}}}\left(1 - C\right),
\end{equation}
where $\gamma_{\rm{re}}=\gamma_A/\gamma_B$, $\gamma_A$ and $\gamma_B$ are the  surface tensions of pure A and B components, respectively. 

In this model, we use the Onsager principle to determine the time evolution of the contact radius $\dot R$ and the contact angle $\dot \theta$ by minimizing the Rayleighian defined by~\cite{Doi2013}
\begin{eqnarray}  \label{ray}
\begin{aligned}
  \Ray=\dot F+\Phi,
\end{aligned}
\end{eqnarray}
where $\dot F$ is the time change rate of the free energy of the system, and $\Phi$ is the energy dissipation function.

The change rate of free energy $\dot F$ has two parts: the interface free energy contribution $\dot F_C$
and the Marangoni flow contribution $\dot F_M$. We assume that the droplet size is less than the capillary length and the droplet is nearly flat $| h'|\ll 1$, then the sum of the interfacial energy is written
as
\begin{eqnarray}  \label{free_energu20}
\begin{aligned}
   F_C=& \int_0^R 2 \pi r  \left[ \gamma(r) \sqrt{ 1 +  h'^2(r)} -  \gamma_{A}\cos\theta_{eA}   \right]dr\\
   = &\int_0^R 2 \pi r \left[ \gamma(r) \left( 1 + \frac{1}{2} h'^2(r) \right ) -  \gamma_{A}\left(1-\frac{1}{2}\theta_{eA}^2\right)  \right]dr.
\end{aligned}
\end{eqnarray}
where  the surface tension $\gamma$ of solution depends on the composition in the solution, and $\theta_{eA}$  is the equilibrium contact angle for A component, $\gamma_{A}\cos\theta_{eA}=\gamma_{SV}-\gamma_{LS}$. Here $\gamma_{SV}$ and $\gamma_{LS}$ are the surface tension of the substrate/vapor and the liquid/substrate interfaces, respectively. Therefore, $\dot F_C$ has the form
\begin{eqnarray}  \label{dotfree_energu20}
\begin{aligned}
   \dot F_C=& \int_0^R 2 \pi r  \left[ \gamma(r)h'(r) \dot h'(r)+\dot C(r) \frac{\partial\gamma(r)}{\partial C(r)} \left( 1 + \frac{h'^2(r)}{2}  \right ) \right]dr\\+&2\pi R\dot R\left[\gamma(R)\left( 1 + \frac{h'^2(R)}{2}  \right )-  \gamma_{A}\left(1-\frac{1}{2}\theta_{eA}^2\right)  \right],
   \end{aligned}
\end{eqnarray}
while the Marangoni flow contribution to the change rate of free energy~\cite{Man2017},
\begin{equation}\label{marangonidot1}
      \dot{F}_M=-\int_0^R 2\pi r \bm {v}_s \frac{\partial \bm\gamma}{\partial r}dr,
\end{equation}
where $\bm v_s$ is the fluid velocity at the liquid-vapor interface.

We use the lubrication approximation to calculate the dissipation function. Therefore, the energy dissipation $\Phi$  taking place in the system is
\begin{equation}\label{phi1}
       \Phi=\frac{\eta}{2}\int_0^R\int_0^h 2\pi r \left(\frac{\partial\bm u}{\partial z}\right)^2dzdr.
\end{equation}
where $\eta$ is the viscosity of the fluid, and $\bm u$ is the fluid velocity inside the droplet as shown in Fig.~2.

According to the lubrication theory, the general expression of the fluid velocity $\bm u$ as a
function of $z$ is
\begin{equation}\label{u1}
       \bm u(r,z,t)=\left(Az^2+Bz+C\right)\bm i.
\end{equation}
The boundary conditions of $\bm u$ are,
\begin{equation}\label{boundary}
\begin{aligned}
       \bm u(r,0,t)=&0,\\
       \bm u(r,h,t)=&\bm v_s(r,t),
\end{aligned}
\end{equation}
where $\bm v_s (r, t)$ is the fluid velocity at the liquid-vapor interface. To simplify the calculation
of $\Phi$, we define a height averaged fluid velocity as
\begin{equation}\label{vf1}
\begin{aligned}
       \bm v_f(r,t)=\frac{1}{h}\int_0^h \bm u(r,z,t)dz.
\end{aligned}
\end{equation}
The definitions of $\bm v_f$ and $\bm v_s$ are schematically shown in Fig.~\ref{fig:drop2}, $\bm v_f (r)$ is the height averaged fluid velocity at position $r$. Then, $\bm u(r, z, t)$becomes
\begin{equation}\label{u2}
\begin{aligned}
      \bm  u=\frac{3\left(\bm v_s-2\bm v_f\right)}{h^2}z^2+\frac{2\left(3\bm v_f-\bm v_s\right)}{h}z.
\end{aligned}
\end{equation}

Combining Eqs.~(\ref{phi1}) and ~(\ref{u2}), the  energy dissipation caused by the fluid flow inside the droplet becomes
\begin{equation}\label{phi2}
       \Phi=\int_0^R 2\pi r \frac{\eta}{2h}\left[12\left(\bm v_f-\frac{\bm v_s}{2}\right)^2+\bm v_s^2\right]dr,
\end{equation}

Minimizing the Rayleighian, $\Ray=\Phi+\dot{F}$, with respect to $\bm v_s(r)$ leading to
\begin{equation}\label{vs}
 \bm v_s=  \frac{3}{2}\bm v_f +\frac{h}{4\eta}\frac{\partial\bm \gamma}{\partial r }.
\end{equation}

The velocity $v_f$ and the time change rate of concentration $\dot C$ are obtained from the mass conservation equation. The liquid volume conservation equation is written as
\begin{equation} \label{mass1}
   \dot h = - \frac{1}{r} \frac{\partial (rv_f h)}{\partial r}  - J.
\end{equation}
Since $h(r,t)$ is given by Eq.~(\ref{eqn:height}), $\dot h$ is expressed as a function of
$\dot R$ and $\dot H$. Therefore, by integrating Eq.~(\ref{mass1}), the height averaged fluid velocity $v_f(r)$ has a form
\begin{equation} \label{vf2}
  v_f=r\frac{\dot R}{R}-\left(\frac{r}{2V}+\frac{r^3}{2\pi R^4h}\right)\dot V-\frac{1}{rh}\int_0^r r'Jdr'.
\end{equation}
The conservation equation for the volatile component is written as
\begin{equation} \label{mass2}
   \frac{\partial{\left(Ch\right)}}{\partial t} = - \frac{1}{r} \frac{\partial (rv_f Ch)}{\partial r}-J,
 \end{equation}
where we have ignore the diffusion of each component. Combining Eqs.~(\ref{mass1}) and (\ref{mass2}), we obtain the evolution equation of the  volatile component  concentration,
\begin{equation}   \label{mass3}
   \frac{\partial C}{\partial t} = - v_f \frac{\partial C}{\partial r} - \frac{J}{h}\left(1-C\right).
\end{equation}
Inserting Eqs.~(\ref{eqn:height}), (\ref{eqn:volume}), (\ref{eqn:angle}), and (\ref{eqn:surface_tension2}) into Eq.~(\ref{dotfree_energu20}), the time change rate of the interfacial free energy $\dot F_C$ becomes
\begin{eqnarray}  \label{dotfree_energu21}
\begin{aligned}
   \dot F_C=& \int_0^R 2 \pi r^3 \left[\left(\gamma_A-\gamma_B\right)C+\gamma_B\right]\left(\frac{16V}{\pi^2R^8}\dot V-\frac{64V^2}{\pi ^2R^9}\dot R\right)dr-2\pi R\dot R\gamma_A\left(1-\frac{1}{2}\theta_{eA}^2\right) \\
   +&\int_0^R 2 \pi r \dot C \left(\gamma_A-\gamma_B\right)\left(1+\frac{r^2\theta^2}{2R^2}\right)dr+2\pi R\dot R\left[\left(\gamma_A-\gamma_B\right)C(R)+\gamma_B\right]\left( 1 + \frac{\theta^2}{2}  \right ).
 \end{aligned}
\end{eqnarray}
Inserting the expression of $v_s$ and $v_f$ into Eq.~(\ref{marangonidot1}), $\dot F_M$ is calculated as
\begin{equation}\label{marangonidot2}
\begin{aligned}
      \dot{F}_M=&-\int_0^R 3\pi r^2\left(\gamma_A-\gamma_B\right)\frac{\partial C}{\partial r}\left[\frac{\dot R}{R}-\left(\frac{1}{2V}+\frac{r^2}{2\pi R^4h}\right)\dot V-\frac{1}{2\pi r^2h}\int_0^r r'Jdr'\right]dr\\
      -&\int_0^R\frac{\pi h r}{2\eta}\left(\gamma_A-\gamma_B\right)^2\left(\frac{\partial C}{\partial r}\right)^2dr.
\end{aligned}
\end{equation}
Moreover, inserting the expression of $v_s$ and $v_f$ into Eq.~(\ref{phi2}) , $\Phi$ is calculated as
\begin{equation}\label{phi3}
       \Phi=\int_0^R  \frac{\pi \eta r}{h}\left[3r^2\left(\frac{\dot R}{R}-\frac{\dot V}{2V}-\frac{r^2\dot V}{2\pi R^4h}-\frac{1}{r^2h}\int_0^r r'Jdr'\right)^2+\frac{h^2\left(\gamma_A-\gamma_B\right)^2}{4\eta^2}\left(\frac{\partial C}{\partial r}\right)^2\right]dr.
\end{equation}

The Onsager principle states that $\dot R$ is determined by the condition $\partial\Ray/ \partial\dot R=0$, and $\Ray=\dot F+\Phi$. Using Eq.~(\ref{dotfree_energu21}), $\partial\dot F_C/ \partial\dot R$ is writing as follows,
\begin{eqnarray}  \label{dotfree_energu31}
\begin{aligned}
 \frac{\partial \dot F_C}{\partial\dot R}=&-\int_0^R\frac{8\pi\theta^2r^3}{R^3} \left[\left(\gamma_A-\gamma_B\right)C+\gamma_B\right]dr
 +\int_0^R 2 \pi r \frac{\partial\dot C }{\partial\dot R}\left(\gamma_A-\gamma_B\right)\left(1+\frac{r^2\theta^2}{2R^2}\right)dr \\ +&2\pi R\left[\left(\gamma_A-\gamma_B\right)C(R)+\gamma_B\right]\left( 1 + \frac{\theta^2}{2}  \right )-2\pi R\gamma_A\left(1-\frac{1}{2}\theta_{eA}^2\right).
 \end{aligned}
\end{eqnarray}
By Eqs.~(\ref{vf2}) and (\ref{mass3}), we have the following expression of $\partial\dot C/ \partial\dot R$,
\begin{eqnarray}  \label{partial_dotC}
\begin{aligned}
 \frac{\partial\dot C}{ \partial\dot R}=-\frac{r}{R}\frac{\partial C}{ \partial r}.
 \end{aligned}
\end{eqnarray}
Inserting Eq.~(\ref{partial_dotC}) into Eq.~(\ref{dotfree_energu31}), $\partial\dot F_C/ \partial\dot R$ has the form,
\begin{eqnarray}  \label{dotfree_energu41}
\begin{aligned}
 \frac{\partial \dot F_C}{\partial\dot R}=&-\int_0^R \frac{8\pi\theta^2r^3}{R^3}\left(\gamma_A-\gamma_B\right)Cdr
 -\int_0^R\frac{2 \pi r^2}{R}\frac{\partial C}{ \partial r}\left(\gamma_A-\gamma_B\right)\left(1+\frac{r^2\theta^2}{2R^2}\right)dr \\ -&2\pi\gamma_B\theta^2R+2\pi R\left[\left(\gamma_A-\gamma_B\right)C(R)+\gamma_B\right]\left( 1 + \frac{\theta^2}{2}  \right )-2\pi R\gamma_A\left(1-\frac{1}{2}\theta_{eA}^2\right).
 \end{aligned}
\end{eqnarray}
Using integration by parts, this can be rewritten as,
\begin{eqnarray}  \label{dotfree_energu51}
\begin{aligned}
 \frac{\partial \dot F_C}{\partial\dot R}
  =&\int_0^R 4 \pi \left(\gamma_A-\gamma_B\right) \frac{Cr}{R}\left(1-\frac{r^2\theta^2}{R^2}\right)dr-2\pi R\left(\gamma_A -\gamma_B\right)-\pi R\left(\gamma_B \theta^2-\gamma_A\theta_{eA}^2\right).
 \end{aligned}
\end{eqnarray}
Then the expression of $\partial\dot F_M/ \partial\dot R$ can be obtained from  Eq.~(\ref{marangonidot2}),
\begin{equation}\label{marangonidot3}
\begin{aligned}
      \frac{\partial \dot F_M}{\partial\dot R}=&-\int_0^R\frac{3\pi r^2}{R}\left(\gamma_A-\gamma_B\right)\frac{\partial C}{\partial r}dr\\
      =&\int_0^R 6\pi \left(\gamma_A-\gamma_B\right)\frac{Cr}{R}dr-3\pi \left(\gamma_A-\gamma_B\right)RC(R).
\end{aligned}
\end{equation}
 Similarly, $\partial\Phi/ \partial\dot R$ is obtained from Eq.~(\ref{phi3}),
\begin{equation}\label{phi4}
\begin{aligned}
      \frac{\partial \Phi}{\partial\dot R} =&\int_0^R  \frac{6\pi \eta r^3}{hR}\left(\frac{\dot R}{R}-\frac{\dot V}{2V}-\frac{r^2\dot V}{2\pi R^4h}-\frac{1}{r^2h}\int_0^r r'Jdr'\right)dr\\
      =&\int_0^R  \frac{6\pi \eta r^3}{hR}\left(\frac{\dot R}{R}-\frac{\dot V}{2V}-\frac{r^2\dot V}{2\pi R^4h}+\frac{\dot V}{2\pi r^2h}+\frac{1}{r^2h}\int_r^R r'Jdr'\right)dr.
\end{aligned}
\end{equation}
Inserting Eqs.~(\ref{eqn:height}), (\ref{eqn:j}) and (\ref{eqn:ja}) into Eq.~(\ref{phi4}),   $\partial\Phi/ \partial\dot R$ is calculated as,
\begin{equation}\label{phi5}
\begin{aligned}
      \frac{\partial \Phi}{\partial\dot R} =\frac{3\pi^2\eta \alpha R^4}{2V}\dot R+\frac{3\pi^2\eta R^5\dot V}{8V^2}+\frac{3\pi \eta\theta_0R_0^2}{2\tau_{\rm{ev}}R^2}\int_0^R\frac{ r}{h^2}\left[\int_r^R r'\left(C-RH\right)dr'\right]dr,
\end{aligned}
\end{equation}
where $ \alpha=\ln(R/2\epsilon)- 1$ is a parameter which is regarded as constant in the subsequent analysis~\cite{Man2016}.

Combining Eqs.~(\ref{ray}), (\ref{dotfree_energu51}), (\ref{marangonidot3}), and (\ref{phi5}), we have the evolution equations of the droplet
\begin{eqnarray}\label{eqn:rdot}
\begin{aligned}
\tau_{\rm{ev}}\dot R=&\frac{\left(\gamma_{\rm{re}}-1\right)\theta V_0^\frac{1}{3}}{3\gamma_{\rm{re}}\alpha k_{\rm{ev}}}
\left[\int_0^R C\frac{r}{R^2}
\left(\frac{2r^2\theta^2}{R^2}-5\right)dr+\frac{\theta^2-\gamma_{\rm{re}}\theta_{eA}^2}{2\left(\gamma_{\rm{re}}-1\right)}+
\frac{3}{2}C\left(R\right)+1\right]
\\-&\frac{R\dot V\tau_{\rm{ev}}}{4\alpha V}-\frac{\theta \theta_0R_0^2}{4\alpha R^3}
\int_0^R \frac{r}{h^2}\left[\int_r^Rr'\left(C-RH\right)dr'\right]dr,
\end{aligned}
\end{eqnarray}
where $ k_{\rm{ev}}=\tau_{\rm{re}}/\tau_{\rm{ev}}$ is the evaporation rate parameter, the character relaxation time $\tau_{\rm{re}}$ is defined by $\tau_{\rm{re}}=\eta V_0^\frac{1}{3}/\gamma_A$. 

Since $\dot\theta$ is related to $\dot V$ by (see Eq.~(\ref{eqn:angle})),
\begin{eqnarray}\label{eqn:thetadot1}
\dot \theta=\theta\frac{\dot V}{V}-3\theta\frac{\dot R}{R}.
\end{eqnarray}
 Eq.~(\ref{eqn:rdot}) gives the following time evolution equation for $\theta$,
\begin{eqnarray}\label{eqn:thetadot2}
\begin{aligned}
\tau_{\rm{ev}}\dot \theta=&\frac{\left(\gamma_{\rm{re}}-1\right)\theta^2 V_0^\frac{1}{3}}{\gamma_{\rm{re}}\alpha k_{\rm{ev}}R}
\left[\int_0^R C\frac{r}{R^2}\left(5-\frac{2r^2\theta^2}{R^2}\right)dr
-\frac{\theta^2-\gamma_{\rm{re}}\theta_{eA}^2}{2\left(\gamma_{\rm{re}}-1\right)}
-\frac{3}{2}C\left(R\right)-1\right]\\
+&\frac{\theta\dot V\tau_{\rm{ev}}}{V}\left(1+\frac{3}{4\alpha }\right)
+\frac{3\theta^2 \theta_0R_0^2}{4\alpha R^4}\int_0^R \frac{r}{h^2}\left[\int_r^Rr'\left(C-RH\right)dr'\right]dr.
\end{aligned}
\end{eqnarray}

It is worth noting that when C is set to 1, Eqs.~(\ref{eqn:rdot}) and (\ref{eqn:thetadot2}) reduce to the model of single component droplet cases. When $C_0=1$,  Eq.~(\ref{eqn:rdot}) can be reduced to
\begin{eqnarray}\label{eqn:rdot2}
\begin{aligned}
\tau_{\rm{ev}}\dot R=&\frac{\left(\gamma_{\rm{re}}-1\right)\theta V_0^\frac{1}{3}}{3\gamma_{\rm{re}}\alpha k_{\rm{ev}}}
\left[\frac{\theta^2-5}{2}+\frac{\theta^2-\gamma_{\rm{re}}\theta_{eA}^2}{2\left(\gamma_{\rm{re}}-1\right)}+\frac{5}{2}\right]
-\frac{R\dot V\tau_{\rm{ev}}}{4\alpha V}\\-&\frac{\theta \theta_0R_0^2}{8\alpha R^3 }\left(1-RH\right)
\int_0^R \frac{r\left(R^2-r^2\right)}{h^2}dr.
\end{aligned}
\end{eqnarray}
Inserting the parabolic form of $h(r,t)$ into Eq.~(\ref{eqn:rdot2}) and integrating the last term, we have
\begin{eqnarray}\label{eqn:rdot3}
\begin{aligned}
\tau_{\rm{ev}}\dot R=\frac{\theta V_0^\frac{1}{3}}{6\alpha k_{\rm{ev}}}
\left(\theta^2-\theta_{eA}^2\right)
-\frac{R\dot V\tau_{\rm{ev}}}{4\alpha V}-\frac{\theta \theta_0R_0^2R}{16\alpha  H^2}\left(1-RH\right)\left(\alpha+1\right).
\end{aligned}
\end{eqnarray}
The last term of Eq.~(\ref{eqn:rdot3}) can be written as a linear function of $J_A$ by using the definition in Eq.~(\ref{eqn:ja}),
\begin{eqnarray}\label{eqn:rdot4}
\begin{aligned}
\tau_{\rm{ev}}\dot R=\frac{\theta V_0^\frac{1}{3}}{6\alpha k_{\rm{ev}}}
\left(\theta^2-\theta_{eA}^2\right)
-\frac{R\dot V\tau_{\rm{ev}}}{4\alpha V}-\frac{\tau_{\rm{ev}}\theta R^2J_A}{4\alpha  H^2}\left(1-RH\right)\left(\alpha+1\right).
\end{aligned}
\end{eqnarray}
Inserting Eqs.~(\ref{eqn:vdot}) and (\ref{eqn:j}) into Eq.~(\ref{eqn:rdot4}),  then the last term of Eq.~(\ref{eqn:rdot4}) can be written as a linear function of $\dot V$
\begin{eqnarray}\label{eqn:rdot5}
\begin{aligned}
\tau_{\rm{ev}}\dot R=\frac{\theta V_0^\frac{1}{3}}{6\alpha k_{\rm{ev}}}
\left(\theta^2-\theta_{eA}^2\right)
-\frac{R\dot V\tau_{\rm{ev}}}{4\alpha V}+\frac{\tau_{\rm{ev}}\theta\dot V}{4\pi\alpha  H^2}\left(\alpha+1\right).
\end{aligned}
\end{eqnarray}
Based on Eqs.~(\ref{eqn:volume}) and (\ref{eqn:angle}), the volume of droplet can be written as, $V=\pi H^2R/\theta$. Then inserting such form of $V$ into Eq.~(\ref{eqn:rdot5}), the equation of $\dot R(t)$ for a drying single-component droplet is obtained as
\begin{eqnarray}\label{eqn:rdot6}
\begin{aligned}
\tau_{\rm{ev}}\dot R=\frac{\theta V_0^\frac{1}{3}}{6\alpha k_{\rm{ev}}}
\left(\theta^2-\theta_{eA}^2\right)+\frac{\tau_{ev}R\dot V}{4 V},
\end{aligned}
\end{eqnarray}
which is consistent with the previous evolution equation of contact radius $R$ for evaporating pure droplets~\cite{Man2016}. Inserting Eq.~(\ref{eqn:rdot6}) into (\ref{eqn:thetadot1}),  the time evolution equation of the droplet contact angle, $\dot\theta$, for a drying single-component droplet is written as  
\begin{eqnarray}\label{eqn:thetadot3}
\tau_{ev}\dot \theta=\frac{\theta^2 V_0^\frac{1}{3}}{2\alpha k_{\rm{ev}}R}
\left(\theta_{eA}^2-\theta^2\right)+\frac{\tau_{ev}\theta\dot V}{4V}.
\end{eqnarray}

\bigskip
\subsection{Comparison with experimental results when $\gamma_{\rm re}<1$}

Our model is also suitable for the case of $\gamma_{\rm re}<1$. As an extension of the paper, we discuss the shape evolution of a drying binary droplet when $\gamma_{\rm re}<1$. Based on the Young's equation, the relationship between the equilibrium contact angle of pure A and B droplet, $\theta_{eA}$ and $\theta_{eB}$, can be written as, $\cos\theta_{eB}=\gamma_{\rm re}\cos\theta_{eA}$. In order to make the model more convenient to use in the case of $\gamma_{\rm re}<1$, we replace $\theta_{eA}$ by  $\theta_{eB}$ in the evolution equations. Also, for comparing with experiments~\cite{Williams2021}, we assume both of the two components can evaporate and  introduce a relative evaporation rate parameter into the present theory model, $J_{\rm re}=J_A/J_B$, where $J_B$ is the evaporation rate for pure B droplet in dry ambient air, $RH_B=0$.
Then the evaporation rate becomes,
\begin{equation}\label{eqn:j2}
     J(r)=J_{A}\left[C(r)-RH+\frac{1-C}{J_{\rm re}}\right].
\end{equation}
With the modified definition of $J$ in Eq.~(\ref{eqn:j}), the volume change rate $\dot{V}$ becomes
\begin{equation} \label{eqn:vdot2}
  \dot V(t)=-\int_0^R 2\pi rJ_{A}\left[C(r)-RH+\frac{1-C}{J_{\rm re}}\right] dr,
\end{equation}
and the height averaged velocity $v_f(r)$ has a new form
\begin{equation} \label{vf3}
  v_f=r\frac{\dot R}{R}-\left(\frac{r}{2V}+\frac{r^3}{2\pi R^4h}\right)\dot V-\frac{1}{rh}\int_0^r J_{A}r'\left[C(r)-RH+\frac{1-C}{J_{\rm re}}\right]dr'.
\end{equation}
Then the time change rate of the concentration of volatile component $\dot C(r)$ becomes,
\begin{equation}   \label{massc2}
  \dot C = - v_f \frac{\partial C}{\partial r} - \frac{J_{A}}{J_{\rm re}h}\left[J_{\rm re}\left(C(r)-RH\right)-C\right]\left[1-C(r)\right].
\end{equation}
 Then, inserting the new $\dot V$, $\dot C$, and $v_f\left(r\right)$ into the dissipation function $\Phi$, and the time change rate of free energy, $\dot F$, results in an new Rayleighian function. Then, we repeat the same minimization process to the Rayleighian and replace $\theta_{eA}$ by $\theta_{eB}$, leading to an new evolution equation of the contact radius, $R(t)$, 
\begin{eqnarray}\label{eqn:rdot7}
\begin{aligned}
\tau_{\rm{ev}}\dot R=&\frac{\left(\gamma_{\rm{re}}-1\right)\theta V_0^\frac{1}{3}}{3\gamma_{\rm{re}}\alpha k_{\rm{ev}}}
\left[\int_0^R C\frac{r}{R^2}
\left(\frac{2r^2\theta^2}{R^2}-5\right)dr+\frac{\theta^2-\theta_{eB}^2}{2\left(\gamma_{\rm{re}}-1\right)}+
\frac{3}{2}C\left(R\right)\right]
-\frac{R\dot V\tau_{\rm{ev}}}{4\alpha V}\\-&\frac{\theta \theta_0R_0^2}{4\alpha R^3}
\int_0^R \frac{r}{h^2}\left[\int_r^Rr'\left(C+\frac{1-C}{J_{\rm re}}-RH\right)dr'\right]dr,
\end{aligned}
\end{eqnarray}
The evolution equation of the contact angle $\theta(t)$ becomes 
\begin{eqnarray}\label{eqn:thetadot4}
\begin{aligned}
\tau_{\rm{ev}}\dot \theta=&\frac{\left(\gamma_{\rm{re}}-1\right)\theta^2 V_0^\frac{1}{3}}{\gamma_{\rm{re}}\alpha k_{\rm{ev}}R}
\left[\int_0^R C\frac{r}{R^2}\left(5-\frac{2r^2\theta^2}{R^2}\right)dr
-\frac{\theta^2-\theta_{eB}^2}{2\left(\gamma_{\rm{re}}-1\right)}
-\frac{3}{2}C\left(R\right)\right]\\
+&\frac{\theta\dot V\tau_{\rm{ev}}}{V}\left(1+\frac{3}{4\alpha }\right)
+\frac{3\theta^2 \theta_0R_0^2}{4\alpha R^4}\int_0^R \frac{r}{h^2}\left[\int_r^Rr'\left(C+\frac{1-C}{J_{\rm re}}-RH\right)dr'\right]dr.
\end{aligned}
\end{eqnarray}

\begin{figure}
\begin{center}
\includegraphics[bb=458 678 0 847, scale=1.0, draft=false]{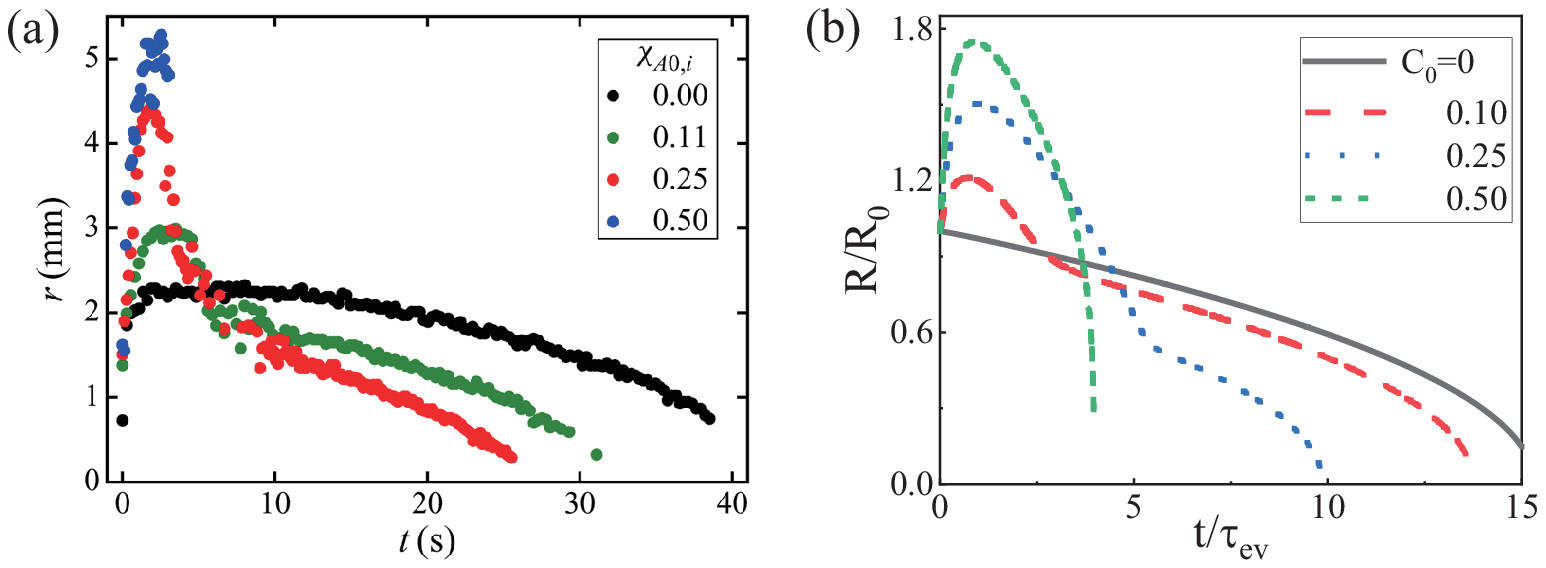}
\caption{(a)  Evolution of droplet contact radius, $r$, of an
evaporating ethanol/water droplet on heated substrate ($70^{\circ}C$) for various initial concentration of ethanol, $\chi_{A0,i}$. Reprinted with the permission from Reference \cite{Williams2021}. Copyright (2020) Cambridge University Press. (b) Corresponding theoretical results calculated by our theory model. $R(t)/R_0$ is the droplet contact radius, and $C_0$ is the initial concentration for the volatile A component, which is consistent with $\chi_{A0,i}$ in (a). All calculations are done for $\theta_0=0.4$, $RH=0$, $\theta_e^B=0.4$, and $\epsilon=10^{-7}$. All the other parameters used for the calculation are given in the experimental parameters and theoretical framework of Reference~\cite{Williams2021}. }
\label{fig:com}
\end{center}
\end{figure}

Williams et al.~\cite{Williams2021} studied the evaporation of ethanol/water droplet on a high energy substrate, and found that increasing the initial ethanol concentration ($\chi _{A0,i}$) can enhance the droplet spreading, resulting in a larger maximum contact radius and shorter overall droplet lifetime, as shown in Figure~\ref{fig:com}(a). The enhanced spreading of droplet radius is explained by the fact that when $\chi _{A0,i}$ increases, the initial liquid/vapor surface tension decreases, and also the surface tension gradient from the apex to the contact line increases. Both effects enhance the spreading of contact line. 

In order to compare with this finding, we carefully set values of parameters used in our model according to real experimental values used in~\cite{Williams2021}. We calculated the total evaporation rate, $\dot V_0$, of a pure water droplet by using the data in Figure~(11) of \cite{Williams2021}. Then, we obtained the characteristic evaporation time of water droplet, $\tau_{\rm{ev}}^B\approx0.24 s$ ($\tau_{\rm{ev}}=-V_0/\dot V_0$). According to Eqs. (2.12) and (2.13) in \cite{Williams2021}, we obtained the ratio of evaporation rate between ethanol ($J_A$) and pure water ($J_B$) as $J_{\rm{re}}=J_A/J_B\approx10.26$. Combing the two values of $\tau_{\rm ev}$ and $J_{\rm re}$, we obtained the characteristic evaporation time of ethanol, $\tau_{\rm{ev}}^A=\tau_{\rm{ev}}^B/10.26\approx0.023 s$, which was used as the scale time in our model. In addition, Table 1 and Eq.~(2.2) of \cite{Williams2021} give the values of the surface tension of ethanol ($\gamma_A$) and water ($\gamma_B$), which are $\gamma_A\approx2.24\times10^{-2} Nm^{-1}$ and $\gamma_B\approx6.61\times10^{-2} Nm^{-1}$, respectively. Then, the surface tension ratio of the faster evaporation component over the slower one is $\gamma_{\rm{re}}\approx0.34$.  Finally, with the values of the initial volume $V_0$, viscosity $\eta_A$, $\gamma_A$ and $\tau^A_{\rm{ev}}$, the evaporation rate parameter $k_{\rm{ev}}=\eta_A V_0^{1/3}/(\gamma_A \tau_{\rm{ev}}^A)$ of our present model has a value $k_{\rm{ev}}\approx1.88\times10^{-3}$. The calculated time dependent contact radius from our model is shown in Figure~\ref{fig:com} (b) for various initial concentration of the more volatile component, $C_0$. It is clear that when $C_0$ increases from $0$ to $0.3$, the droplet spreading is enhanced which is qualitatively consistent with the experimental results. The main reason of this phenomenon is the same as the explanations in \cite{Williams2021}, which is due to the enhanced Marangoni effects induced by the inhomogeneous surface tension.

\bigskip
\subsection{Effects of the evaporation rate $k_{\rm ev}$}
\begin{figure}[h!]
\begin{center}
\includegraphics[bb=0 0 504 264, scale=0.9, draft=false]{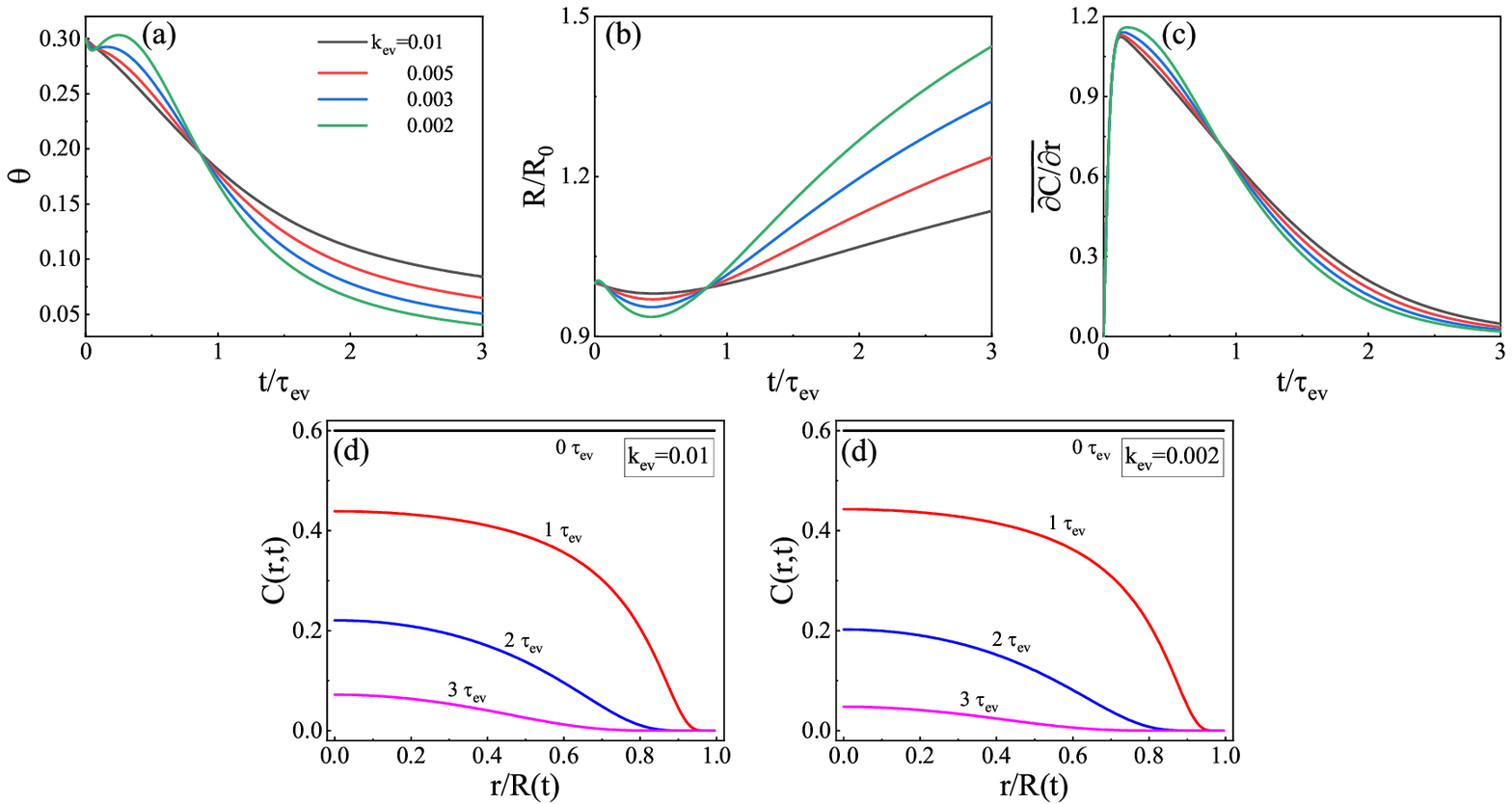}
\caption{ Evolution of (a) droplet contact angle $\theta(t)$, (b) droplet contact radius $R(t)/R_0$, and (c)  averaged concentration gradient $\overline{\partial C/\partial r}$ of an
evaporating binary aqueous solution droplet for various values of $k_{\rm ev}$. The evolution of the distribution of volatile component concentration within the droplet for (d) $k_{\rm ev}=0.01$, and (e) $k_{\rm ev}=0.002$.  All other parameters are $RH=0.0$, and $C_0=0.6$. }
\label{fig:kev}
\end{center}
\end{figure}

In our model, the evaporation rate is determined by both $k_{\rm{ev}}$ and $RH$. The larger value of $k_{\rm{ev}}$ or smaller value of $RH$ represent the faster evaporation rate. Figure~\ref{fig:kev} (a) shows the effect of $k_{\rm{ev}}$
on the time variation of the contact angle $\theta(t)$, and Figure~\ref{fig:kev}(b) is the corresponding time evolution of the contact radius $R(t)$. When the evaporation rate
is relatively large ($k_{\rm{ev}}=0.01$), the contact angle
decreases monotonically in time. When $k_{\rm{ev}}$ decreases to $0.005$,
$\theta(t)$ starts to have a maximum value (i.e. a peak
contact angle $\theta_{\rm{pk}}$ appears).  The maximum value increases
when  $k_{\rm{ev}}$ decreases.  Meanwhile, the contact radius $R(t)$ shows
the decreasing-increasing transition as $k_{\rm{ev}}$ decrease as it is shown in
Figure~\ref{fig:kev}(b).
An important point here is that $\overline{\partial C/\partial r}$  is almost
independent of $k_{\rm{ev}}$: $\overline{\partial C/\partial r}$  is determined by how much of
the volatile component has evaporated, and it  is independent how fast the volatile component
has evaporated. Figure~\ref{fig:kev}(d) and (e) shows that $C(r,t)$ at the edge quickly goes to zero when $RH=0$, while the change of $C(r,t)$ in the droplet center is relative slow. As the average concentration gradient is mainly determined by the difference between the value of $C(r,t)$ in the droplet center and in the edge, the Marangoni flow is not strongly affected by $k_{\rm{ev}}$.
Therefore the change of $\theta(t)$ shown in Figure~\ref{fig:kev}(a) is due to the
effect of evaporation: evaporation decreases the contact angle as it is indicated by
the last two terms on the right hand side of Eq.~(\ref{eqn:thetadot2}).
\bigskip
\subsection{The values used for parameters}

 The justification for the values used for parameters in our calculations are provided as follows. 
 \begin{itemize}
 \item[(1)] The evaporation rate parameter $k_{\rm{ev}}$.
  
 $k_{\rm{ev}}$ is defined by a ratio of two characteristic times, $k_{\rm{ev}}=\tau_{re}/\tau_{ev}$, where $\tau_{re}=\eta V_0^{1/3}/\gamma_A$ and $\tau_{\rm{ev}}=V_0/\left|\dot V_0\right|$. The time $\tau_{ev}$ represents the characteristic evaporation time for pure A droplet (of initial size $V_0$), and $\tau_{re}$ represents the relaxation time: the time needed for the pure A droplet (initially having contact angle $\theta_0$) to have the equilibrium contact angle $\theta_e$. The parameters $\eta$ and $\gamma_A$  are viscosity and surface tension of pure A droplet, respectively. Due to the large difference between these parameters in different systems and conditions, the value range of $k_{\rm{ev}}$ is relatively large, and $k_{\rm{ev}}\in[10^{-9},10]$ in general.
 
  \item[(2)] The relative surface tension $\gamma_{\rm re}$.
  
Using the droplets made of  propylene glycol (PG) and water, Cira et al.~\cite{Cira2015} observed the $\theta_{pk}$ phenomenon. The surface tension of  water and PG in the experiment correspond to $\gamma_A=73 ~\rm{mN~m^{-1}}$ and $\gamma_B=36 ~\rm{mN~m^{-1}}$, respectively~\cite{Cira2015}. Put these parameters into $\gamma_{\rm{re}}=\gamma_A/\gamma_B$, we have  $\gamma_{\rm{re}}\approx2.03$. Thus we set $\gamma_{\rm re}=2$ in our calculations.
  
 \item[(3)]  The initial contact angle $\theta_0$.
 
The error between $\cos\theta$ and $1-\theta^2/2$ is within $0.5\%$ when $\theta<\pi/6$. Moreover, the initial contact angle usually is set in between $0-0.5$ in many other studies using lubrication approximation theory. For example, Williams et al.~\cite{Williams2021} set the initial aspect ratio $H_0/R_0=0.2$, which is corresponds to $\theta_0\approx0.4$, where the evaporation of ethanol/water droplets was studied both experimentally and theoretically. Thus it is reasonable to set the initial contact angle $\theta_0=0.3$ in our calculations.
\end{itemize}